\documentclass[conference]{IEEEtran}
\IEEEoverridecommandlockouts
\usepackage{cite}
\usepackage{amsmath,amssymb,amsfonts}
\usepackage{algorithmic}
\usepackage{graphicx}
\usepackage{textcomp}
\usepackage{xcolor}

\def\BibTeX{{\rm B\kern-.05em{\sc i\kern-.025em b}\kern-.08em
    T\kern-.1667em\lower.7ex\hbox{E}\kern-.125emX}}
\begin{document}

\title{A distributed system perspective on Backscatter systems}

\author{
\IEEEauthorblockN{Guan Jincheng}
\IEEEauthorblockA{\textit{University of Science and Technology of China} \\
\textit{guanjincheng@mail.ustc.edu.cn}}
\and
\IEEEauthorblockN{Zhang Jun}
\IEEEauthorblockA{\textit{University of Science and Technology of China} \\
\textit{zhangjun03@mail.ustc.edu.cn}}
}

\maketitle

\begin{abstract}

Backscatter system is a system based on backscatter communication technology, which is a low cost, low power consumption and easy to deploy communication technology. At present, the backscatter technology is mainly applied to RFID tags and the Internet of Things and other fields. With the rapid development of the Internet of Things, the application of backscatter systems is increasing. Moreover, the backscatter system is essentially a distributed system, but existing research rarely conducts studies and analyses from a distributed perspective. This paper conducts a study on the backscattering system from the perspective of distributed systems, comprehensively reviewing the basic principles of the backscattering system, and analyzing the distributed system architectures of different backscattering systems. Then, it introduces the application scenarios, research status and challenges of the backscattering system, and finally discusses the future research directions of the backscattering system, hoping to provide references for future research.
\end{abstract}


\section{Introduction}

With the rapid development of technologies such as the Internet of Things (IoT), smart cities, and ubiquitous computing, the scale and quantity of communication nodes have increased dramatically, presenting unprecedented challenges to the power consumption, cost, and scalability of communication technologies. Traditional active wireless communication technologies (such as Wi-Fi, Bluetooth, ZigBee, etc.) often have issues like excessive power consumption and high deployment costs when deployed on a large scale, making it difficult to meet the requirements of next-generation large-scale distributed applications. In this context, backscatter communication technology, as an emerging low-power and low-cost wireless communication method, has increasingly attracted attention from both the academic and industrial communities.

The core of backscatter communication is that the equipment does not actively emit radio signals but instead modulates and reflects the existing radio frequency (RF) signals (such as Wi-Fi, FM radio, TV broadcast, etc.) in the environment to achieve information transmission. This technology was first widely applied in RFID (Radio Frequency Identification) systems for short-distance communication between tags and readers. In recent years, with the introduction of new technologies such as ambient backscatter and Wi-Fi backscatter, the application scenarios of backscatter communication have become more extensive. Currently, backscatter systems are also widely used in various fields such as environmental sensing, smart home, and healthcare.

One important feature of the backscatter communication system is that the power consumption of the nodes is extremely low. The nodes can even operate without batteries. Moreover, there are a large number of nodes and they are highly dispersed. These are typical characteristics of a distributed system, and these features make backscatter particularly suitable for large-scale distributed applications. Since the backscatter nodes usually operate in environments with severely limited resources, their communication and coordination must rely on distributed algorithms and protocols. However, current research on backscatter communication lacks studies from the perspective of distributed systems.

Distributed systems are an important field in computer science research. The characteristics of distributed systems can be applied to many system architectures. The existing theories of distributed systems are also very mature. Applying the theories and technologies of distributed systems to the backscattering system can solve problems such as reliability, interference management, delay and synchronization in the backscattering network. Moreover, it can also significantly improve the performance of the backscattering system. Therefore, analyzing the backscattering technology from the perspective of distributed systems is a meaningful task.

This paper conducts a systematic analysis and summary of the principles, system architecture, application status, key challenges, and future research directions of backscatter communication from the perspective of distributed systems. Finally, our contributions are as follows:

\begin{enumerate}
    \item A survey of the fundamental principles of backscatter communication systems
    \item A survey of the distributed‐systems architecture of backscatter systems, including communication protocols, data management and processing, and resource classification
    \item An overview of the application scenarios for backscatter systems
    \item An analysis of research progress and key challenges for backscatter systems from a distributed‐systems perspective
    \item The important future research directions of the backscattering system
\end{enumerate}

The structure of this paper is as follows: Section 2 surveys the fundamental principles of backscatter communication systems; Section 3 presents an overview of backscatter system architecture; Section 4 reviews typical applications of backscatter systems; Section 5 analyzes the key challenges in this field from a distributed‐systems perspective; Section 6 discusses future research directions; and Section 7 concludes the paper.

\section{Fundamental Principles of Backscatter Systems}

Backscatter communication is a special wireless communication method. The node itself does not actively generate radio signals but instead modulates and reflects existing ambient RF signals to achieve ultra‐low‐power information transmission \cite{liu2013ambient}. Specifically, a backscatter node embeds information into the reflected signal by periodically toggling the impedance state of its antenna (e.g., changing the load impedance) to modulate the incident RF wave. When the receiver captures these modulated reflected signals, it can demodulate them to extract the carried information\cite{ensworth2017ble}.

A typical backscatter system generally consists of the following three main components:
\begin{enumerate}
\item RF source: a device that generates the carrier signal, which can be a dedicated RF source (e.g., an RFID reader) or an ambient RF source (e.g., a Wi-Fi router, FM broadcast, or television signal).
\item backscatter node: It does not require its own transmitter. It only needs to adjust the ground impedance of its own antenna and use the carrier signal for communication. The energy consumption is extremely low, and it can even achieve battery-free operation.
\item Receiver: a device that dedicated to receiving and demodulating the reflected signal (e.g., an RFID reader, a specialized receiver, or a Wi-Fi device).
\end{enumerate}

Specifically, a typical backscatter communication process can be divided into the following steps:
\begin{enumerate}
\item Generation and propagation of the RF signal. The backscatter system requires an RF signal source that produces a continuous or periodic RF carrier (for example, the signal emitted by a dedicated RFID reader or by a Wi-Fi router).
\item Reflection and modulation by the node. The backscatter node does not actively transmit an RF signal; instead, it changes its antenna load impedance to reflect and modulate the incoming RF signal. Internal circuitry toggles between different impedance states via a load switch, altering the strength of the antenna’s reflected signal and thus embedding data into the reflected RF wave. Because the node only switches its load impedance without generating any RF energy, its power consumption remains extremely low.
\item Reception and demodulation. A dedicated receiving device (such as an RFID reader or a wireless receiver in an ambient backscatter system) captures the reflected, modulated signal and extracts the embedded information using appropriate demodulation algorithms. The receiver typically has high sensitivity and can distinguish between the direct-path carrier and the modulated reflection to recover the data.
\item Signal encoding and transmission performance optimization. To enhance the performance of the backscatter system, researchers have proposed various modulation and encoding schemes, as well as corresponding performance optimization methods.
\end{enumerate}

Due to the aforementioned technical design, backscatter communication inherently possesses several advantages. For instance, it features ultra-low power consumption, low cost, and simple hardware implementation. As a result, it is widely applied in fields such as the Internet of Things, sensor networks, smart homes, and medical monitoring.\cite{xu2018practical}.

From an implementation perspective, backscatter communication can be classified into three main categories:
\begin{enumerate}
\item Traditional RFID backscatter: communication between a dedicated reader and RFID tags, suitable for applications such as logistics tracking and inventory management.
\item Ambient backscatter: leveraging existing ambient RF signals (e.g., television, Wi-Fi) to enable communication without requiring a dedicated signal source for the node\cite{lu2018ambient}.
\item Wi-Fi backscatter: specifically utilizing Wi-Fi signals to realize backscatter communication, widely used in indoor IoT scenarios\cite{kellogg2016passive}.
\end{enumerate}

In recent years, ambient backscatter and Wi-Fi backscatter have attracted particular attention. Research hotspots include modulation scheme optimization, coding scheme design, signal demodulation algorithm development, and interference and noise mitigation\cite{zhang2016hitchhike}. These advances have greatly expanded the application scope of backscatter communication. This enables the backscatter technology to evolve from the traditional RFID technology to a broader field of the Internet of Things.

\section{Distributed-Systems Architecture of Backscatter Systems}

Backscatter communication technology is a typical low-power wireless communication solution, and its system design has inherent distributed characteristics. Since the backscatter network is usually composed of a large number of nodes, and the nodes communicate with each other in a passive or semi-passive manner, this structure has the typical characteristics of a distributed system. For example, high scalability, decentralized management, and heterogeneity of node resources. Therefore, researchers and users need to understand the distributed system architecture of the backscatter system in order to effectively manage and optimize the system. Specifically, we can analyze its architecture from its communication protocol design, data management and processing mechanisms, as well as the classification and scheduling of resources.

\subsection{Design of Communication Protocols}
In a backscatter communication network, nodes exchange information by sharing a wireless channel. Therefore, suitable distributed communication protocols must be adopted, including:
\begin{enumerate}
\item Distributed Medium Access Control (MAC) Protocols. The passive or semi‐passive nature of backscatter nodes requires the protocol to be highly streamlined and low‐overhead. Specific methods include:
\begin{itemize}
\item Carrier Sense Multiple Access (CSMA). Nodes listen to ambient signals before transmitting to avoid collisions. For example, the Backscatter‐CSMA protocol allows nodes to sense the carrier and determine an appropriate transmission opportunity\cite{lu2018ambient}.
\item Random Access and Slotted ALOHA Protocols. Nodes randomly select time slots to send data, statistically reducing collisions; for instance, the ALOHA protocol used in ambient backscatter\cite{bharadia2015backfi}.
\item Distributed implementations of TDMA and FDMA protocols. Nodes coordinate in a distributed manner to share communication resources—such as time slots or frequency bands—through negotiation and consensus, achieving efficient resource allocation and utilization.
\item Protocols specifically designed for large‐scale backscatter networks. That effectively coordinate multi‐node communication, improving throughput and network scalability\cite{hessar2019netscatter}.
\end{itemize}
\item Distributed Routing and Multi‐Hop Communication Protocols. Because the communication range between nodes is usually limited, multi‐hop communication and distributed routing protocols can extend network coverage.
\begin{itemize}
    \item Random propagation mechanisms based on flooding and gossip: Nodes randomly select neighboring nodes to forward data, which improving information dissemination efficiency and achieving robust data propagation\cite{peng2018plora}.
    \item Adaptive routing algorithms based on energy consumption or link quality. Enabling self‐organizing, adaptive routing decisions among nodes.
\end{itemize}

\item Distributed Spectrum Management and Interference Control. Since in high‐density node deployment environments, interference becomes prominent.
\begin{itemize}
    \item Distributed Channel Selection: Nodes autonomously choose suitable frequencies or channels to reduce interference and collisions\cite{zhang2016hitchhike}.
    \item Interference‐Aware Adaptive Backscatter Modulation Schemes: Nodes negotiate and perform real‐time sensing to mitigate interference effects\cite{parks2014turbocharging}.
\end{itemize}
\end{enumerate}

\subsection{Data Management and Processing Mechanisms}

In a distributed backscatter communication system, node resources are typically constrained, so efficient data management is required within the distributed system.
\begin{enumerate}
\item Distributed Data Aggregation
\begin{itemize}
\item Nodes preprocess and compress data locally, then transmit via backscatter to edge devices or gateways.
\item Edge devices integrate, aggregate, and process the dispersed data. And reducing the total network communication load\cite{xu2018practical}.
\end{itemize}
\item Data Caching and Distributed Storage
\begin{itemize}
\item Implement distributed data caching on relay or semi‐passive nodes. This can avoid redundant transmissions and reduce network communication overhead.
\item Distributed caching strategies can effectively alleviate storage pressure on individual nodes. This can enhance overall data reliability and availability.
\end{itemize}
\item Edge Computing–Assisted Processing
\begin{itemize}
\item Utilize a small number of resource‐rich edge nodes to perform data analysis, feature extraction, anomaly detection, and other tasks.
\item Edge computing nodes form a complementary structure with backscatter nodes, effectively reducing the computational load on the central server\cite{kellogg2016passive}.
\end{itemize}
\end{enumerate}

\subsection{Resource Classification and Management}

Based on the heterogeneity and functional differences of the reverse scattering node resources, the distributed system can be classified into the following categories:
\begin{enumerate}
\item Energy Resource Classification
\begin{itemize}
\item Passive Nodes: Rely entirely on ambient RF energy, have no battery, and perform extremely limited communication tasks\cite{liu2013ambient}.
\item Semi‐Passive Nodes: Equipped with a small battery or capacitor, capable of handling moderate data caching or processing tasks\cite{song2021advances}.
\item Active Nodes: Have stable power support and stronger computation and communication capabilities, undertaking more complex tasks such as edge nodes or gateways\cite{xu2018practical}.
\end{itemize}
\item Functional Role Classification
\begin{itemize}
\item Sensing Nodes: Responsible for environmental sensing, data collection, and simple processing\cite{yang2018cooperative}.
\item Relay Nodes: Used for data forwarding and network coverage extension.
\item Edge Nodes: Perform data aggregation, analysis, and management decision functions.
\end{itemize}
\item Communication Role Classification
\begin{itemize}
\item End‐User Nodes: Primarily collect and transmit data, bearing a relatively light communication load\cite{yang2018cooperative}.
\item Relay Nodes: Handle multi‐hop data relaying and extend transmission paths.
\item Gateway Nodes: Core nodes that connect the backscatter network to external systems, responsible for network control and data uploading\cite{kellogg2016passive}.
\end{itemize}
\end{enumerate}

\subsection{Typical Architecture Examples}

A typical distributed backscatter network architecture often adopts three- or four-layer structures:
\begin{itemize}
\item End‐User Node Tier: A large number of low‐power sensing nodes.
\item Relay Node Tier: Nodes that extend network coverage and enhance connectivity.
\item Edge Node Tier: Data aggregation and preliminary processing nodes, reducing network traffic.
\item Cloud Server Tier: A centralized data analysis and storage center providing advanced intelligent services\cite{xu2018practical}\cite{kellogg2016passive}.
\end{itemize}
This layered architecture can efficiently address network scalability and resource constraints.

Below are several typical backscatter communication system architecture examples, covering a range of designs from traditional to modern:
\begin{enumerate}
  \item \textbf{Single‐Reader Architecture.} In this arrangement, the RF source and the reader are built into one unit. This device continuously emits a carrier wave while simultaneously listening for the modulated reflection from the Backscatter Device (BD). Its primary benefit lies in the simplicity of system design and ease of deployment; however, because the signal must travel from and return to the same location, it suffers double path loss, which restricts communication range. Most conventional RFID systems employ this configuration\cite{song2021advances}.

  \item \textbf{Dual‐Base Architecture.} Here, the RF source and the reader are physically separated and deployed at distinct locations. The RF source broadcasts a continuous‐wave signal; the BD modulates and reflects that signal, and the reader picks up and demodulates the return. By optimally positioning the RF source and the reader, one can reduce path loss and extend range. The trade-off is added deployment complexity, since careful coordination and timing synchronization between the source and reader are required. The LoRea system exemplifies this approach to achieve long-range backscatter links\cite{varshney2017lorea}.

  \item \textbf{Ambient Backscatter Architecture.} This topology leverages preexisting ambient RF transmissions (such as TV broadcasts or Wi-Fi) as carriers. The BD imposes its own modulation onto those signals, enabling communication without a dedicated RF transmitter. Its advantage is very low cost and power consumption, but performance hinges entirely on the presence and stability of ambient RF signals. Research has shown that this architecture can support battery-free low-power communication\cite{liu2013ambient}.

  \item \textbf{LoRa Backscatter Architecture.} In this scheme, low-power BDs generate LoRa-compatible waveforms using backscatter techniques, combining backscatter’s minimal energy use with LoRa’s inherently long reach. This makes it well-suited for IoT deployments requiring both extended range and tiny power budgets. On the downside, system design becomes more intricate, since precise frequency control and specialized modulation circuits are necessary. Experimental results demonstrate that LoRa backscatter can cover several hundred meters in real-world settings\cite{talla2017lora}.

  \item \textbf{Wireless Power Transfer Backscatter Network Architecture.} In this model, dedicated energy transmitters wirelessly deliver power to BDs, which then communicate via backscatter. By enabling self-powered, ultra-low-power devices, this architecture is attractive for large-scale IoT rollouts. The drawback is that energy transmitters must be installed alongside the communication infrastructure, raising both cost and complexity. It is particularly suitable for scenarios such as smart cities or industrial monitoring, where a vast number of distributed devices need minimal maintenance\cite{han2017wirelessly}.
\end{enumerate}

\section{Applications of Distributed Backscatter Systems}
With the rapid advancement of wireless communication technologies and sensor networks, backscatter communication is increasingly emerging as a fundamental component for building distributed sensing systems, due to its advantages of low power consumption, low cost, and the absence of active radios. Distributed backscatter systems demonstrate significant application potential across several key areas, such as environmental monitoring, industrial IoT, smart cities, and healthcare\cite{liu2013ambient,jiang2023backscatter,lin2017survey}. The following sections elaborate on the application prospects and technical advantages of these systems in four representative scenarios.

\subsection{Distributed Sensing for Environmental Monitoring}
Wireless sensors based on backscatter mechanisms offer an efficient and scalable solution for environmental monitoring. Such systems can be deployed in the wild, on farmlands, or in other ecological areas to enable real-time collection of environmental parameters such as canopy temperature and moisture content\cite{salvati2023emerging}. In particular, optical fiber sensors, as a special type of backscatter medium, have been widely applied in Distributed Acoustic Sensing (DAS) and Distributed Temperature Sensing (DTS) systems due to their ability to maintain high sensitivity over long distances. DAS can detect geological dynamics such as seismic waves, landslides, and underground activity\cite{markomsystematic}, while DTS technology enables continuous temperature monitoring along the fiber path and is widely used in scenarios like climate change studies, forest fire warning systems, and water resource management\cite{failleau2016development}. Overall, fiber-optic-based distributed backscatter systems offer both spatial continuity and real-time temporal responsiveness, along with high robustness and long-range coverage, making them an essential component of future environmental monitoring systems.

\subsection{Scalable Solutions for Industrial IoT}
Backscatter communication is regarded as a highly promising low-power communication technology in the Industrial Internet of Things (IIoT), particularly suitable for energy-constrained and maintenance-intensive industrial environments\cite{huang2023resources}. Traditional industrial systems often face challenges such as high power consumption, complex wiring, and high maintenance costs when deploying large-scale sensor networks. In contrast, distributed wireless backscatter systems support battery-free operation and leverage ambient signal reflections for communication, making them an ideal medium for industrial automation and real-time monitoring. Furthermore, integrating backscatter systems with Multiple-Input Multiple-Output (MIMO) technology can significantly enhance network throughput and communication reliability\cite{wang2023distributed}. Thanks to their flexible deployment, low power consumption, and long-term operational capability, these systems can be widely applied in scenarios such as production line status monitoring, equipment health management, and environmental safety surveillance, providing a sustainable technological foundation for IIoT development.

\subsection{Smart City Deployments with Distributed Backscatter Infrastructure}
As urban informatization continues to advance, the construction of efficient, low-power, and widely covered sensing and communication networks has become a core requirement for smart city development. Distributed backscatter communication technology, with its low cost and non-intrusive deployment characteristics, provides a novel approach to implementing city-scale sensor networks. For example, backscatter communication has been applied in typical smart city scenarios such as urban traffic flow monitoring and waste management tracking\cite{gupta2023hybrid}. In addition, ambient backscatter communication can utilize existing FM broadcast and other ambient wireless signals to perform communication tasks without requiring additional infrastructure\cite{liu2019next}, significantly reducing deployment costs and enhancing system availability. By combining with distributed acoustic sensing technology, backscatter systems can also be used to identify vehicle operating conditions, monitor crowd activity, and detect urban noise\cite{cho2023ai}. Integrating distributed backscatter infrastructure into urban communication and sensing systems can help achieve intelligent and fine-grained city management while ensuring energy efficiency and system stability.

\subsection{Distributed Backscatter Systems for Healthcare Applications}
In the healthcare sector, especially in applications with stringent energy efficiency and device miniaturization requirements, backscatter communication technology also demonstrates great potential. Its low power consumption and low interference make it particularly suitable for constructing wearable sensor networks, implantable medical devices, and other miniature embedded systems\cite{talla2017lora}. Wearable devices integrated with backscatter modules can continuously collect users’ vital signs—such as heart rate, body temperature, and activity intensity—enabling round-the-clock health monitoring without frequent charging or maintenance. Additionally, the deployment of wide-area backscatter systems in hospitals and rehabilitation centers enables real-time remote monitoring of patient status by medical staff, effectively improving the quality and efficiency of healthcare services. Due to the extremely small size and ultra-low power consumption of backscatter devices, they offer significant advantages in non-invasive monitoring, telemedicine, and elderly care, and are expected to become a key technological enabler for personalized and intelligent healthcare in the future.

\section{Security and Key Challenges of Distributed Backscatter Systems}
As a low-power and scalable architecture for communication and sensing, distributed backscatter systems have recently attracted extensive attention across various fields. However, to achieve efficient and stable operation in large-scale practical scenarios, numerous technical and security challenges must be addressed. This section presents an in-depth discussion of the security mechanisms and key technical bottlenecks in current research.

\subsection{Security and Privacy Protocol Design}
\subsubsection{Lightweight Encryption Mechanisms}
Due to the limited computational resources of backscatter devices, traditional high-strength encryption algorithms are not applicable\cite{jin2024enhanced}. Current research focuses on developing lightweight encryption protocols tailored for low-power devices, such as hash-based authentication mechanisms, simplified block ciphers, and hardware-friendly symmetric key algorithms\cite{rana2023enhancing}. Additionally, strategies like key pre-distribution and partial key agreement can effectively reduce key management overhead and enhance overall system security.

\subsubsection{Distributed Key Management and Authentication}
In large-scale backscatter networks, centralized key management is often insufficient in terms of efficiency and reliability. Distributed key management schemes can improve the system's scalability and robustness while reducing the risk of single points of failure\cite{rana2023enhancing}. Moreover, lightweight authentication mechanisms based on challenge-response or one-time passwords can be used for device registration, authentication, and message integrity verification, ensuring secure communication in heterogeneous environments.

\subsubsection{Physical Layer Security Mechanisms}
Physical layer security (PLS) leverages the uncertainty and dynamics of wireless channels to enhance the security of passive device communications. For instance, artificial noise injection can effectively disrupt unauthorized receivers\cite{hou2020physical}, while tag selection and signal direction control techniques can improve the secrecy capacity of communications\cite{kaveh2023secrecy}. Combining these with reconfigurable intelligent surfaces (RIS) enables directional transmission to legitimate users and provides effective resistance against eavesdropping\cite{abideen2025advancements}.

\subsubsection{Privacy Protection Mechanisms}
In sensitive applications such as healthcare and environmental monitoring, data privacy is of particular concern. Covert communication techniques can reduce detectability, thereby concealing communication behaviors and avoiding malicious interception\cite{gu2023act}. In addition, privacy-aware data aggregation mechanisms can prevent individual device data from being tracked or exposed while maintaining system efficiency\cite{agnihotri2024energy}. These technologies provide a foundation for building user-friendly and regulation-compliant distributed backscatter networks.

\subsection{Key Technical Challenges}
\subsubsection{Synchronization and Coordination in Large-Scale Deployments}
The core of efficient communication in large-scale distributed backscatter networks lies in the precise synchronization between the incident signal and reflection nodes\cite{dunna2021syncscatter}. This typically requires symbol-level or even sub-symbol-level timing precision. However, hardware inconsistencies among nodes, signal propagation delays, and dynamic environmental changes can significantly affect synchronization accuracy, making it difficult to maintain stable synchronization during real-world operation\cite{hessar2019netscatter}. Although some distributed coding mechanisms, such as random access and hash-based coding, can mitigate collisions caused by concurrent transmissions, they still rely on fundamental synchronization support. Achieving efficient and robust synchronization under resource constraints remains a core design challenge.

\subsubsection{Interference Management in Dense Networks}
As network scale and device density increase, interference becomes a major issue. Interference in backscatter systems mainly originates from ambient RF signals, non-orthogonal transmissions from other backscatter nodes, and direct link interference (DLI) from carrier emitters, all of which can overwhelm reflected signals\cite{xu2018practical}. To address this, researchers have proposed solutions such as successive interference cancellation (SIC), orthogonal frequency division multiplexing (OFDM), and frequency shifting. In multi-reader systems, interference can also be reduced through frequency and spatial division\cite{biswas2021direct}. Building a distributed backscatter network with strong interference resistance and adaptability to varying device densities is fundamental to improving reliability and scalability.

\subsubsection{Trade-off Between Scalability and Network Capacity}
The communication range of traditional backscatter systems is limited, constraining their applicability in large-scale and wide-area scenarios. Single-reader systems face bottlenecks in synchronization and throughput when scaling\cite{patel2024analyzing}. To overcome this, researchers have proposed multi-hop backscatter networks and distributed coding mechanisms to expand coverage and support concurrent access by more devices\cite{zhao2020towards,hessar2019netscatter}. However, maintaining high capacity while improving scalability still requires innovative network architectures and protocol support, especially under resource-constrained conditions where trade-offs between node density, throughput, and reliability must be considered.

\subsubsection{Reliability and Fault Tolerance Design}
Due to the inherently weak and interference-prone signals of backscatter communication, links are often unstable and less reliable\cite{wang2012efficient}. In distributed systems, single points of failure can significantly affect overall performance, necessitating fault-tolerant mechanisms to ensure continuous operation. Techniques such as redundant deployment, data replication, and functional backups can improve survivability in cases of node failure or communication disruption\cite{ledmi2018fault}. In critical and high-reliability applications such as industrial control and medical monitoring, systems must possess self-recovery and fault isolation capabilities.

\subsubsection{Data Consistency and Integrity in Distributed Environments}
In backscatter networks composed of numerous resource-constrained devices, ensuring the consistency and integrity of sensed data during transmission, aggregation, and storage is another major challenge\cite{aldin2019consistency}. Data aggregation mechanisms must minimize redundancy and communication load while preventing information loss or distortion\cite{agnihotri2024energy}. Additionally, during concurrent data writing or updating by multiple nodes, conflicts may occur, requiring version control, conflict detection, and merging strategies to maintain consistency.

\subsubsection{Resource Management and Allocation Optimization}
Devices in distributed backscatter systems are typically limited by energy, spectrum bandwidth, and the precision of reflection coefficient control\cite{ye2022resource}. Efficient dynamic allocation of these resources while maintaining communication performance is a key challenge in system design. Current research has explored the use of distributed scheduling algorithms, power control protocols, and adaptive reflection coefficient mechanisms to improve resource utilization\cite{dunna2021syncscatter}. Furthermore, game theory and machine learning approaches have been introduced to address optimization problems in resource allocation, enabling robust and efficient system operation under varying loads.

\section{Future Research Directions}
Distributed backscatter systems, as a low-power and large-scale deployable communication technology, have garnered widespread attention in both academia and industry in recent years. Despite some breakthrough progress, many challenges remain unresolved. To further advance this field, future research can focus on the following key directions:

\subsection{Deep Integration of Backscatter Communication with Emerging Wireless Technologies (e.g., 5G/6G)}
With the commercialization of 5G and the accelerated research on 6G, integrating backscatter communication with these next-generation wireless networks is becoming an important approach to enhance network coverage, connection density, and energy efficiency:
\begin{itemize}
    \item \textbf{Massive low-power connectivity}: Backscatter communication features extremely low power consumption\cite{kaplan2023direct}. When combined with the high connection density capability of 5G/6G networks, it can achieve truly ubiquitous “Internet of Everything” connectivity.
    \item \textbf{Combination with Non-Orthogonal Multiple Access (NOMA)}: NOMA allows multiple devices to share the same time and spectrum resources simultaneously\cite{das2024adaptive}. Integrating NOMA with backscatter communication holds the potential to realize higher spectral efficiency and lower energy consumption in 6G environments.
    \item \textbf{Introduction of Reconfigurable Intelligent Surfaces (RIS)}: As a key technology for 6G, RIS can dynamically adjust signal propagation paths to enhance the strength and directionality of backscatter signals\cite{abideen2025advancements}, thereby improving communication quality and physical layer security.
\end{itemize}
Future research should focus on designing efficient protocols and architectures to achieve seamless integration of backscatter technology with 5G/6G networks and promote its deployment in large-scale intelligent terminal access scenarios.

\subsection{Optimization and Innovation of Distributed Algorithms and Protocols}
In distributed backscatter networks, traditional centralized communication protocols struggle to meet scalability and robustness requirements. Therefore, designing efficient distributed algorithms and protocols is a critical research topic:
\begin{itemize}
    \item \textbf{MAC protocol optimization}: Develop distributed medium access control (MAC) protocols suitable for high-density device environments to reduce collisions and improve latency and energy efficiency\cite{cao2019distributed}.
    \item \textbf{AI-driven resource allocation}: Utilize artificial intelligence and reinforcement learning to dynamically optimize strategies for power allocation, spectrum allocation, and time scheduling based on network conditions\cite{ye2022resource}.
    \item \textbf{Robust time synchronization protocols}: Backscatter communication demands high synchronization accuracy\cite{dunna2021syncscatter}. Designing low-power and fault-tolerant time synchronization protocols is fundamental for large-scale deployment.
\end{itemize}
Research in this direction will provide technical support for the stable operation of distributed backscatter systems in dynamic and uncertain environments.

\subsection{Expansion and Validation of Emerging Application Scenarios}
The low power consumption and passive nature of backscatter communication offer broad application potential. Future research should focus on exploring and validating its feasibility and advantages in emerging scenarios:
\begin{itemize}
    \item \textbf{Smart textiles and wearable devices}: Embed communication functionality in fabrics using backscatter technology to realize battery-free smart clothing\cite{chen2024survey}.
    \item \textbf{Augmented reality (AR) and the metaverse}: Integrate backscatter communication modules into AR devices to enable ultra-low-power data feedback and status sensing\cite{chen2024survey}.
    \item \textbf{Underwater communication and extreme environment sensing}: Backscatter technology demonstrates superior energy efficiency in high humidity, high pressure, or electromagnetic shielding environments, making it a promising solution for underwater and post-disaster communications\cite{chen2024survey}.
    \item \textbf{Integrated sensing and communication (ISAC)}: Deeply integrate backscatter with sensing capabilities to simultaneously achieve intelligent environmental perception and low-power communication, applicable in environmental monitoring, healthcare, and other fields\cite{zargari2023sensing}.
\end{itemize}
Exploring new applications not only broadens technological boundaries but also promotes upgrading and expansion of the related industrial chains.

\subsection{Overcoming Key Challenges and Technical Bottlenecks}
Although distributed backscatter communication offers many advantages, it still faces multiple technical bottlenecks and unresolved issues that require ongoing research efforts:
\begin{itemize}
    \item \textbf{Communication range and data rate}: Current systems are limited in coverage and throughput\cite{niu2019overview}. Improvements in coding methods, signal processing techniques, and network architectures are urgently needed to enhance performance.
    \item \textbf{Robustness and interference management}: Ensuring communication quality under multipath fading, device failures, or harsh interference conditions demands the development of more robust modulation, demodulation, and interference resistance mechanisms\cite{xu2018practical}.
    \item \textbf{Security and privacy protection}: For sensitive scenarios such as healthcare and industry, lightweight security mechanisms balancing performance and computational overhead are required to prevent information leakage and data tampering\cite{kaveh2023secrecy}.
    \item \textbf{Standardization and interoperability}: The lack of unified protocols and standards hinders large-scale deployment and compatibility among heterogeneous devices. Future efforts should strengthen research promotion and collaboration within international standard organizations\cite{xu2018practical}.
\end{itemize}
Systematic solutions to these problems will lay a solid foundation for transitioning backscatter communication from laboratory research to practical applications.

In summary, the future development of distributed backscatter systems depends not only on continuous breakthroughs in core technologies but also on deep coordination around application demands and system integration. Through interdisciplinary research, standardization efforts, and industry chain collaboration, it is expected to build a new generation of intelligent, efficient, secure, and scalable low-power communication networks.

\section{Conclusion}
This paper provides a comprehensive survey of backscatter communication technology from the perspective of distributed systems. It begins by introducing the fundamental principles of backscatter communication, highlighting its ultra-low power consumption, low-cost deployment, and typical three-tier architecture consisting of an RF source, backscatter node, and receiver. These characteristics make backscatter communication particularly well-suited for large-scale and resource-constrained Internet of Things (IoT) scenarios. The paper further emphasizes the inherently distributed nature of such systems, which allows for scalable deployment across diverse application domains.

Subsequently, the paper conducts an in-depth analysis of the distributed architecture of backscatter systems, focusing on key aspects such as communication protocol design (including distributed MAC protocols, routing protocols, and interference management strategies), data management mechanisms (such as distributed data aggregation and edge computing), and resource classification approaches (e.g., heterogeneity in energy availability and functional roles of nodes). In this context, the paper identifies critical challenges that arise in distributed backscatter deployments, including synchronization and coordination in large-scale networks, interference management in densely deployed environments, scalability and network capacity limitations, reliability and fault tolerance, data consistency and integrity, as well as security and privacy protection.

In terms of application scenarios, the paper explores the practical potential of distributed backscatter systems in various domains such as environmental monitoring, industrial IoT, smart cities, and healthcare, demonstrating their adaptability and effectiveness in real-world settings. It also reviews recent research progress in addressing existing challenges through the development of distributed algorithms and frameworks, including time synchronization techniques, resource allocation schemes, and robust security models tailored for the unique constraints of backscatter systems.

Furthermore, the paper outlines future research directions aimed at advancing this technology, such as integration with emerging wireless infrastructures like 5G and 6G, innovation in distributed algorithm design, validation in novel and dynamic scenarios, and overcoming existing technical bottlenecks. Finally, it underscores the importance of interdisciplinary collaboration and the establishment of industry standards to facilitate the transition of backscatter communication systems from experimental prototypes to practical, scalable, and secure deployments in the next generation of wireless communication and IoT ecosystems.

\bibliographystyle{plain}
\bibliography{reference}

\end{document}